\documentclass[12pt,letterpaper]{article}	
\usepackage{graphicx,amsmath}
\usepackage{authblk}

\usepackage[style = ieee]{biblatex}
\title{Rewritable Photonic Integrated Circuits Using Dielectric-assisted Phase-change Material Waveguides}

\author[1,2]{Forrest Miller}
\author[1]{Rui Chen}
\author[1,3]{Johannes E. Fr\"och}
\author[3]{Hannah Rarick}
\author[4]{Sarah Geiger}
\author[1,3*]{Arka Majumdar}

\affil[1]{Department of Electrical and Computer Engineering, University of Washington, 185 E Stevens Way NE, Seattle WA}
\affil[2]{Draper Scholar, The Charles Stark Draper Laboratory, 555 Technology Square, Cambridge MA}
\affil[3]{Department of Physics, University of Washington, 3910 15th Ave. NE, Seattle WA}
\affil[4]{The Charles Stark Draper Laboratory, 555 Technology Square, Cambridge MA}

\affil[*]{Corresponding Author: arka@uw.edu}

\begin{document}

\maketitle

\section*{Abstract}
Photonic integrated circuits (PICs) can drastically expand the capabilities of quantum and classical optical information science and engineering. PICs are commonly fabricated using selective material etching, a subtractive process. Thus, the chip's functionality cannot be substantially altered once fabricated. Here, we propose to exploit wide-bandgap non-volatile phase-change materials (PCMs) to create rewritable PICs. A PCM-based PIC can be written using a nano-second pulsed laser without removing any material, akin to rewritable compact disks. The whole circuit can then be erased by heating, and a new circuit can be rewritten. We designed a dielectric-assisted PCM waveguide consisting of a thick dielectric layer on top of a thin layer of wide-bandgap PCMs $Sb_2S_3$ and $Sb_2Se_3$. The low-loss PCMs and our designed waveguides lead to negligible optical loss. Furthermore, we analyzed the spatio-temporal laser pulse shape to write the PICs. Our proposed platform will enable low-cost manufacturing and have a far-reaching impact on the rapid prototyping of PICs, validation of new designs, and photonic education. 

\section{Introduction}
Photonic Integrated Circuits (PICs) are becoming essential for various applications, including optical communication \cite{communicaiton}, sensing \cite{sensor}, and quantum information science \cite{quantum}. While PICs can significantly expand and enhance the performance of these systems, the fabrication methodology for the PICs is complex and expensive: they require high-resolution lithography and etch processes, which must take place in a sophisticated nanofabrication facility \cite{expense}. Additionally, these processes are inherently subtractive, \emph{i.e.}, once fabricated, the wafer cannot be used to fabricate other structures. A low-cost method to fabricate PICs and the ability to rewrite the PIC in the same wafer can help with rapid prototyping.

Chalcogenide-based non-volatile phase change materials (PCMs) provide a promising route to create such rewritable PICs \cite{acsreview, RodgerCLEO, RodgerIEEE}. These PCMs exhibit large changes in their refractive index ($\Delta n>0.5$) when they undergo structural phase transition between the amorphous (aPCM) and crystalline (cPCM) states  \cite{wuttigpcm}. Crystallization can be actuated by holding the PCM above its glass transition temperature ($T_g$) but below the melting temperature ($T_{mp}$) until a crystal lattice forms. Amorphization is achieved by melting and rapidly quenching the PCM. Notably, this micro-structural phase transition is non-volatile, \emph{i.e.}, no external power is required to maintain the state after the material phase is changed \cite{SbSonMZ}. These PCMs have been cycled thousands of times without degradation \cite{delaney} and can potentially be switched for more than $10^{12}$ times \cite{endurance}. Consequently, PCMs are  widely used in rewritable compact disks (CDs) to store information. A writing laser fires pulses to heat segments of the PCMs which, through amorphization or crystallization, write or erase the stored information. The information is then read out using a probing CW laser. However, rewritable CDs are fundamentally different from rewritable PICs. In a CD, the probing light is reflected off the surface, but in PICs, it propagates along the surface. Therefore, to build PICs, PCMs must provide enough refractive index contrast to guide light in-plane while avoiding significant optical loss.

While researchers have already experimentally demonstrated laser-written rewritable meta-optics \cite{pcmmetamaterial, wangNature} in PCM, an expensive femto-second laser was used, and the light did not propagate in a waveguide over a long path. Another work demonstrated only one-way writing of PICs in PCM \cite{laserMachining}, which lacks the rewritable functionality. Some PIC structures have been written in $Ge_2Sb_2Te_5$ (GST) using nano-second lasers \cite{nanosecondPCM}, but the waveguide loss was not considered. 

\begin{figure}[ht]
\centering
\includegraphics[width=4in]{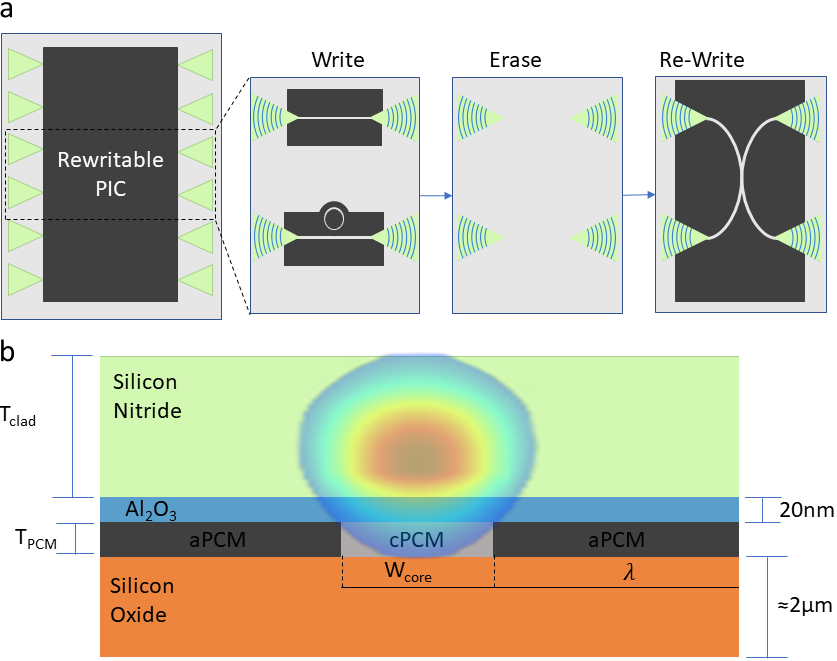}
\caption{a) Schematic of the proposed rewritable PIC platform. A nano-second pulsed laser switches the PCM from the crystalline (light grey) to the amorphous (dark grey) phase. The chip can then be uniformly heated to reset the PCM to the crystalline phase, erasing the written PICs. Subsequently, a different PIC can be written in the same region. b) The proposed dielectric-assisted PCM waveguide geometry for low-loss waveguiding. A simulated guided mode is overlayed, depicting a well-confined mode in the dielectric silicon nitride layer for a low propagation loss. Air clads the upper face of the $Si_3N_4$ layer.}
\label{schematic}
\end{figure}

Here, we present a design of a rewritable PIC based on PCMs. As the probing light is guided in the high-index crystalline PCM, we must ensure near-zero absorptive loss in the PCM. At 1.55 $\mu$m, GST is too lossy with an extinction coefficient of cGST $\kappa_{cGST}\approx1$\cite{jiajiu}. However, wide-bandgap PCMs, such as $Sb_2S_3$ and $Sb_2Se_3$, are suitable thanks to their low absorption \cite{delaney}. These wide-bandgap PCMs also exhibit large enough index contrast between their amorphous and crystalline states ($\Delta n_{SbS} = 0.6$ and $\Delta n_{SbSe}$ = 0.77 at 1.55$\mu$m) \cite{delaney} to confine an optical mode. The loss can be further reduced via a dielectric-assisted PCM structure (Fig. \ref{schematic}b). The propagation loss is estimated at 0.0100 dB/$\mu$m (0.0086 dB/$\mu$m) using $Sb_2S_3$ ($Sb_2Se_3$). We envision that the probing light will be coupled in and out of the chip using pre-fabricated grating couplers, akin to input/output pins in an electronic integrated circuit (Fig. \ref{schematic}a). Finally, we simulate switching dynamics to select the spatio-temporal pulse shape of the writing laser that best achieves a complete and reversible phase transition. Specifically, we show that a nano-second pulsed laser can actuate the phase transition with a spatially Gaussian and temporally rectangular shape. Our proposed rewritable PIC platform could democratize PIC prototyping thanks to its low cost and reusability. This frugal innovation can help with educating students and rapid prototyping of PICs to validate designs.

\section{Low-Loss PCM Waveguides}
While wide-bandgap PCMs have negligible loss in the amorphous state, a non-negligible loss is still present in the crystalline state. We investigated prior works on PCM-integrated ring resonators and estimate the refractive index at $\lambda = 1.55 \mu m$ in the crystalline state to be $n_{cSbS} = 3.33+0.016i$ \cite{ruiarXiv}, and $n_{cSbSe} = 4.23+0.0043i$ \cite{rios,rogergraphene}. We obtained these values by simulating the reported experimental structures and adjusting the extinction coefficient of the PCMs until our simulated loss matched the reported experimental results. The loss in the crystalline state is critical since the probing light is confined in crystalline PCM, which has a higher refractive index than the amorphous phase ($n_{aSbS} = 2.76 +0i$, $n_{aSbSe} = 3.22+0i$) but also higher loss. If the light is perfectly confined in a $cSb_2S_3$ ($cSb_2Se_3$) core, then the unit propagation loss at $\lambda$ becomes $-20\log_{10}[\exp(-\frac{2\pi}{\lambda} \cdot \kappa_{cSbS (cSbSe)})]$, resulting in a unit loss for $cSb_2S_3$ ($cSb_2Se_3$) $\approx 0.56 (0.15) dB/\mu m$. While the loss is overestimated due to the perfect confinement assumption, the result suggests the necessity of a careful waveguide design to achieve a low propagation loss.  

 One straightforward way to reduce the optical loss is to reduce the PCM thickness, which decreases the interaction between the optical mode and the cPCM. Additionally, thick PCMs are generally harder to fully switch with 70nm the thickest to date \cite{GaoIntermediate}. Moreover, thinner PCM layers enable higher switching endurance \cite{RuiBroadband,acsreview}. However, an extremely thin PCM layer does not provide enough index contrast to guide a mode. We mitigate this trade-off by exploiting a dielectric-assisted PCM waveguide architecture \cite{thinSi}, where a thick dielectric layer of $Si_3N_4$ \cite{Kischkat} is deposited on a thin PCM layer (Fig. \ref{schematic}a). This layer may experience small changes in refractive index during writing, but we anticipate such changes will be negligible \cite{SiNtuning}.  The optical mode is mainly confined in the dielectric layer due to the geometry of the PCM layer, mitigating the absorptive loss of cPCMs. Such a waveguide will be written from the chip's “erased” state, where the PCM layer is uniformly crystalline. A PCM waveguide is created by selectively switching the PCM to the amorphous state. We assume a layer of PCM with a thickness $T_{PCM}$ will be deposited on 2 $\mu$m of thermal oxide ($SiO_2$) film on a silicon wafer. A conformal capping material is needed to prevent material reflow and oxidation during switching \cite{NonVolatileReconfig}. Therefore, we encapsulate the PCM with 20nm of atomic layer deposited $Al_2O_3$ and then deposit a $Si_3N_4$ layer with thickness $T_{clad}$ to create the material stack (Fig. \ref{schematic}b).

\begin{figure}[ht]
\centering
\includegraphics[width=4in]{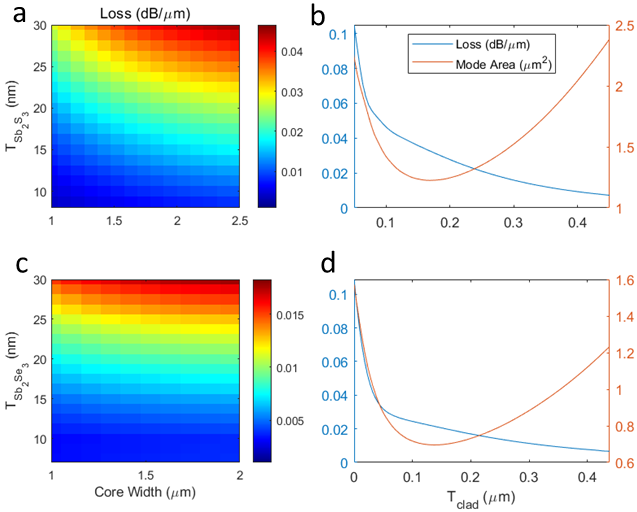}
\caption{a) Waveguide loss (dB/$\mu$m) for a joint parameter sweep of the $Sb_2S_3$ thickness and waveguide core ($cSb_2S_3$) width. b) The loss and effective mode area as a function of the $Si_3N_4$ thickness in the $Sb_2S_3$ design. c) Waveguide loss (dB/$\mu$m) for a joint parameter sweep of the $Sb_2Se_3$ thickness and waveguide core ($cSb_2Se_3$) width. d) The loss and effective mode area as a function of the $Si_3N_4$ thickness in the $Sb_2Se_3$ design.} 
\label{lossPlot}
\end{figure}

We optimize the waveguide geometry for low loss, using the Finite Element Eigenmode (FEM) solver in Ansys Lumerical (Supplement S1). We start by sweeping the $Sb_2S_3$ thickness $T_{PCM}$, and the $cSb_2S_3$ core width $W_{core}$ (Fig. \ref{lossPlot}a). Unsurprisingly, a thinner $T_{PCM}$ yields a lower loss. However, below a thickness of 12 nm, the PCM cannot confine a mode. Since the PCM must guide the mode, we find a compromise between loss and mode confinement at a layer thickness of 15 nm and a core width of 2.0 $\mu$m. The thickness of the $Si_3N_4$ layer is designed around 0.4 $\mu$m as shown in Fig. \ref{lossPlot}b. Our optimized structure exhibits a loss of 0.0100 dB/$\mu$m. 

The loss in the $Sb_2Se_3$ waveguide exhibits much weaker dependence on $W_{core}$ (Fig. \ref{lossPlot}c). This is due to the higher real refractive index of $Sb_2Se_3$, which gives tighter mode confinement. We choose 1.5$\mu$m as the core width to improve integration density. Similar to the $Sb_2S_3$ design, the loss increases with increasing $T_{PCM}$. Here we choose 20 nm as the layer thickness. This $T_{PCM}$ is thin enough to switch and provide low loss while reliably guiding a mode. The $Si_3N_4$ thickness was set to 400 nm to minimize loss while still confining the mode (Fig. \ref{lossPlot}d). This geometry demonstrates a loss of 0.0086 $dB/\mu m$ at 1.55 $\mu m$.

We note that while a PCM thickness of 15nm is sufficient for a straight waveguide, for a bent waveguide, we need slightly thicker PCM and/ or thinner $Si_3N_4$. Based on our simulation, we found a $T_{PCM}$ of 25nm  PCM and a 100nm $T_{clad}$ will be guide light across a 75$\mu$m (25 $\mu$m) bend using $Sb_2S_3$ ($Sb_2Se_3$). These new parameters increase the loss by a factor of 8 (4), but support bending, which is critical for any integrated photonic device. With this architecture, we believe we can optically write directional couplers, Y-splitters and ring/ disk resonators \cite{SiNring}. Components like mode-converters or waveguide crossings rely on small features, and our rewritable PIC may not be suitable to write such small features.

\section{Fixed In/out ports using grating couplers}
The PICs presented here must interface with free-space optics. We propose to accomplish this using fixed grating couplers on the chip, forming optical input/output ports, between which designs can be written, erased, and re-written (Fig. \ref{schematic}a). Edge coupling would also work with this design, but we envision grating couplers enabling easier interfacing. These grating couplers are formed by etching the $Si_3N_4$ layer since the optical modes for both designs are primarily confined in the $Si_3N_4$ layer (Fig. \ref{schematic}b). While this etching step is exactly what we intend to eliminate, we note that after this one etch, thousands of PIC designs can be written and tested on this platform without further etching. It is possible to write grating couplers into the PCM, but such couplers suffer from low coupling efficiencies and must be re-written after each anneal. We optimize these gratings for maximum coupling efficiency using single mode fibers at $25^o$ of incidence using Lumerical's Finite Difference Time Domain (FDTD) simulation (Supplement S2). In the $Sb_2S_3$ design, a grating pitch of 1.01 $\mu$m, a duty cycle of 0.8, and an etch depth of 180 nm resulted in a coupling efficiency of 21\%. In the $Sb_2Se_3$ design, a grating pitch of 0.97 $\mu$m, a duty cycle of 0.56, and an etch depth of 240 nm also resulted in a coupling efficiency of 21\%. The gratings are connected to the waveguides via a tapered section with end width of 4 $\mu$m for both PCMs, which yields a mode overlap of more than 95\% with the guided modes. 
\begin{figure}
\centering
\includegraphics[width = 4in]{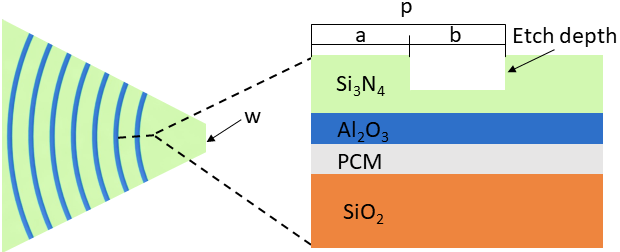}
\caption{A grating coupler and the two-dimensional cross-sectional view along the dashed line. P, the optimal pitch, is 1.01 (0.97) $\mu$m for the $Sb_2S_3$ ($Sb_2Se_3$) design. The optimal duty cycle, $\frac{a}{b}$, is 0.8 (0.56) for the $Sb_2S_3$ ($Sb_2Se_3$) design. The etch depth was 180 (240) nm for $Sb_2S_3$ ($Sb_2Se_3$). In both cases, a taper end width, w, of 4 $\mu$m provides more than 95\% mode overlap between the coupler and the waveguide.}
\label{GC.png}
\end{figure}

\section{Spatio-temporal pulse shaping}
\begin{figure}[!ht]
\centering
\includegraphics[width=4in]{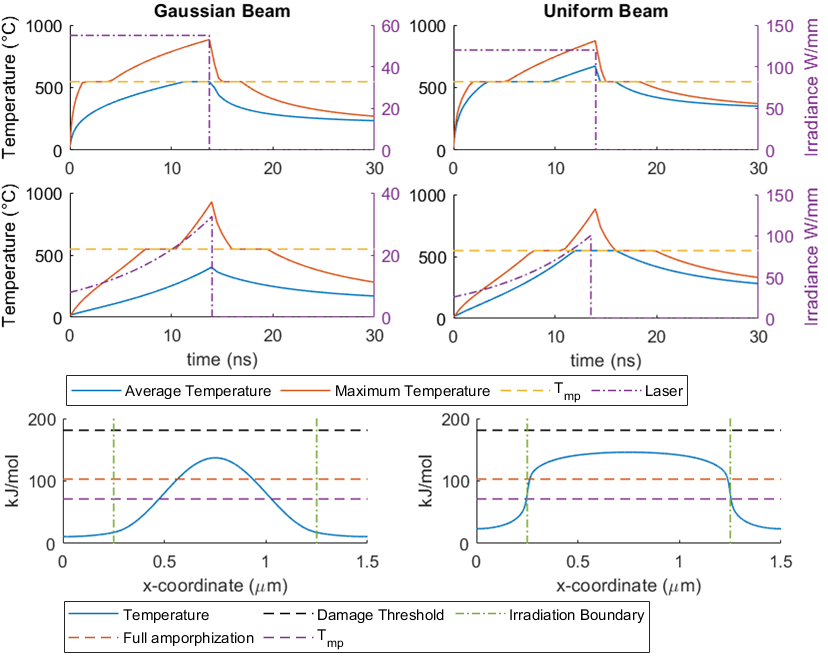}
\caption{a-d) The transient thermal dynamics for different pulses: a) temporally rectangular, spatially Gaussian, b) temporally rectangular, spatially uniform, c) temporally exponentially decaying, spatially Gaussian, d) temporally exponentially decaying, spatially uniform. e, f) Static temperature profile along the spatio-temporal line cut at y = $T_{SbS/2}$ and t = 14 ns for e) a spatially Gaussian beam and f) a spatially uniform beam. The Gaussian beam switches a smaller area of $Sb_2S_3$ than the uniform beam, but the uniform beam has a shorter boundary region (the distance between $T_{mp}$ and full amorphization). Thus a Gaussian beam can write finer features, but a uniform beam will likely have better mode confinement.}
\label{thermalPlot}
\end{figure}

The PICs can be arbitrarily patterned using a pulsed laser and a three-axis translation stage. We simulated four pulse schemes with different spatio-temporal pulse shapes (\ref{thermalPlot}a-d) in COMSOL Multiphysics and select the pulse that best amorphizes the PCM. We expect the ideal shape to be a rectangular spatial shape to provide a smooth boundary region and an increasing temporal shape to produce uniformly switching in the depth direction. However, we found that a Gaussian spatial shape and rectangular temporal shape can switch thin $Sb_2S_3$, offering a simple experimental realization. 450 nm wavelength pulses were delivered directly to the 15 nm film of $cSb_2S_3$ ($n = 4.12 + 1.80i$) in Fig. \ref{schematic}b, so reflection off the alumina and $Si_3N_4$ layers is not considered. The pulses had either a Gaussian or uniform spatial distribution and either a rectangular or exponentially decaying temporal distribution. All pulses last approximately 14 ns and the power was adjusted to achieve a maximum temperature of approximately 900 $^o C$. 

A successful amorphization pulse must satisfy four conditions. First, enough thermal energy must be applied to melt the PCM completely. This requires the delivered energy to heat PCM to its melting temperature and further overcome the latent heat, given by the multiplication of enthalpy of fusion ($H_f$) and mass of the PCM. Second, the cooling rate must exceed $\sim$ 1 $K/ns$ \cite{wuttigpcm}. This ensures the PCM is frozen in the amorphous state \cite{wuttigstorage} and does not undergo unintentional recrystallization. Third, the absorbed thermal energy must not heat the PCM above its boiling point $T_b$, which irreversibly ablates the material. We obtain $Sb_2S_3$ parameters from the literature for the simulations: the melting point is $T_{mp} =$ 547 $^oC$ \cite{delaney}, the enthalpy of fusion is $H_f=$ 47.9 kJ/mol, the boiling point is $T_b=$ 1149 $^oC$, the density is $4.56$ $g/cm^3$ \cite{handbook}, the specific heat is $575.4$ $J/(kg*K)$, and the thermal conductivity is $0.42$ $W/(m*K)$\cite{ruiarXiv}. Material parameters for $Si_3N_4$, $Al_2O_3$, and $SiO_2$ are found from COMSOL's materials library (Supplement S3). Lastly, the boundary region between switched and non-switched regions defines the waveguide edge. This boundary contains material heated to its melting point but not enough to exceed its enthalpy of fusion. This implies a partial amorphization, where a portion of the crystalline structure remains. This is undesirable as a more abrupt index change at the boundary could lead to better mode confinement.

We examine different pulse conditions against these criteria. As shown in Fig. \ref{thermalPlot}a-d, all pulse conditions exceed the melting point and achieve a faster cooling rate than $1 K/ns$. Our hypothesis with temporal modulation was that the exponential ramp would more uniformly heat the PCM layer along its thickness in the vertical direction. This statement is supported by the more linear average temperature curve in \ref{thermalPlot}c,d. However, since the $Sb_2S_3$ layer is very thin, there was little variation in temperature from the top to the bottom of the PCM film. Therefore, we can consider pulses with rectangular temporal modulation. In a thicker PCM layer, we anticipate that temporal modulation of the beam power could enable more uniform heating in the depth direction.

The performance difference between the spatial beam shapes is apparent in the feature size and the boundary region width. A Gaussian mode is better if the desired circuit requires fine features. This is a direct result of the narrower intensity distribution of a Gaussian compared to a uniform beam. With a Gaussian beam, the width of the switched PCM could be smaller than the laser's spot size if the laser power is tuned such that the beam's full-width-half-maximum is lower than the amorphization threshold. A uniform beam lacks this ability to achieve a small feature size. However, its advantage is a narrower boundary region in the PCM. The boundary region for the Gaussian beams traverses a distance of 105 pm (Fig. \ref{thermalPlot}e). In comparison, the boundary is only 27 pm (Fig. \ref{thermalPlot}f) for a spatially uniform beam. This shorter boundary resembles the step-index profiles used in our previous simulations and leads to better mode confinement. Our thermal simulation verifies that the thin PCM layer can be switched entirely with nano-second laser pulses and offers a simple experimental realization with a natural laser beam with a Gaussian spatial and rectangular temporal shape.

\newpage

\section{Conclusion}
In conclusion, we have proposed and numerically verified a rewritable, cost-efficient PIC platform using wide-bandgap PCMs and a cost-efficient nano-second pulse laser. PICs are envisioned to be written (amorphized) by nano-second laser pulses and erased (crystallized) by rapid thermal annealing or even a simple hotplate. We have designed a dielectric-assisted PCM waveguide configuration, allowing low propagation loss for both $Sb_2Se_3$ and $Sb_2S_3$. Efficient grating couplers, working as optical I/O ports, were optimized for both types of waveguides. Comprehensive thermal transfer dynamic simulations were used to verify and select a desirable spatio-temporal pulse condition to ensure complete amorphization with nano-second pulses. This etch-free platform could accelerate PIC fabrication and testing, potentially democratizing PIC fabrication.

\section{Backmatter}
\subsection*{Funding} Research was sponsored by DARPA and the Army Research Office and
was accomplished under Grant Number W911NF-21-1-0368. F.M. is supported by a Draper Scholarship.

\subsection*{Acknowledgments} 
\textbf{Author Contributions} 
A.M. and F.M. conceived the project. F.M. performed simulation. R.C., J.F., and H.R. assisted with simulation. A.M. and S.G. supervised the project. F.M. wrote the manuscript with input from all the authors.

\subsection*{Disclosures} The authors declare no conflicts of interest. 

\subsection*{Data Availability} Data underlying the results presented in this paper are not publicly available at this time but may be obtained from the authors upon request.

% Bibliography
\printbibliography

\end{document}